\documentclass[10pt, conference]{IEEEtran}
\usepackage[utf8]{inputenc}
\usepackage[T1]{fontenc}
\usepackage{fixltx2e}
\usepackage{graphicx}
\usepackage{float}
\usepackage{wrapfig}
\usepackage{rotating}
\usepackage[normalem]{ulem}
\usepackage{amsmath}
\usepackage{textcomp}
\usepackage{marvosym}
\usepackage{wasysym}
\usepackage{amssymb}
\usepackage{hyperref}
\usepackage[margin=2cm]{geometry}
\usepackage{amssymb,amsmath}
\usepackage{float} 
\usepackage{color,soul}
\restylefloat{figure} 
\usepackage{url} 
\usepackage{pdflscape}
\usepackage{setspace}
\usepackage{multicol}
\usepackage{fancybox}
\usepackage{threeparttable} 
\usepackage{listings}
\usepackage{pgfplots}
\usepackage{bchart}
\usepackage{caption}
\usepackage{subcaption}
\usepackage[position=b]{subcaption}
\usepackage{tikz}
\IEEEtriggeratref{17}
\hypersetup{pdftex, colorlinks=true,citecolor=blue,backref=section,urlcolor=gray, linkcolor=blue}

\DeclareCaptionType{eqcap}[][List of equations]

\begin{document}

\title{Performance Localisation \\ 
  {\normalfont\large Brendan Cody-Kenny\IEEEauthorrefmark{1}\IEEEauthorrefmark{2}, Michael O'Neill\IEEEauthorrefmark{1}, Stephen Barrett\IEEEauthorrefmark{2}}\\ [-1ex] }

\author{

  \IEEEauthorblockA{\IEEEauthorrefmark{1}Natural Computing Research \& Applications Group\\
    Michael Smurfit Graduate Business School\\
    University College Dublin\\
    Ireland\\
    \{brendan.cody-kenny, m.oneill\}@ucd.ie}
  \and

  \IEEEauthorblockA{\IEEEauthorrefmark{2}     Distributed Systems Group\\
    School of Computer Science \& Statistics\\
    Trinity College Dublin\\
    Ireland\\
    stephen.barrett@scss.tcd.ie
    }
  
}

\maketitle

\begin{abstract}
Performance becomes an issue particularly when execution cost hinders the functionality of a program. Typically a profiler can be used to find program code execution which represents a large portion of the overall execution cost of a program. Pinpointing where a performance issue exists provides a starting point for tracing cause back through a program.

While profiling shows where a performance issue manifests, we use mutation analysis to show where a performance improvement is likely to exist.
We find that mutation analysis can indicate locations within a program which are highly impactful to the overall execution cost of a program yet are executed relatively infrequently. 
By better locating potential performance improvements in programs we hope to make performance improvement more amenable to automation. 
\end{abstract}


\IEEEpeerreviewmaketitle

\section{Introduction}
\label{sec:introduction}
Software maintenance tasks, such as bug fixing and performance improvement are time consuming \cite{nistor2013discovering}. Once a bug is detected, it may be difficult to understand the existing code and design a fix. This is particularly difficult in larger more complex programs. To aid the diagnosis process, localisation techniques have been developed to highlight what code elements are particularly relevant to a functionality \cite{jones2005empirical} or performance \cite{coppa2012input} defect. Finding where and how an issue manifests in source code can help indicate where a solution is likely to exist.

Improving program performance is frequently of secondary importance to improving functionality when developing software \cite{knuth1974structured}. Focusing developer attention on functionality allows performance issues to manifest. Recent results indicate that improvement is mostly attempted when developers notice a clear improvement opportunity \cite{nistor2013discovering}. Performance improvement is also undertaken to alleviate performance bugs where program execution cost impinges on functional correctness \cite{nistor2013discovering}.  Outside of these prominent scenarios, the implicit nature of how source code results in performance may further allow potential performance opportunities to go unnoticed. Where modern development practices recommend separation of concerns and reuse of code, it becomes increasingly unlikely that developers understand the performance characteristics of the API's and libraries their code depends on \cite{selakovic2016performance}. To aid performance issue detection, static analysis techniques have been developed for locating performance bugs \cite{mudduluru2016efficient} and bottlenecks \cite{conf/issta/ShenLPG15} in code.

Profiling generally refers to measuring the execution cost of a program and is a frequently used technique for finding the location of performance ``bottlenecks'' in code. Profiling can be performed by instrumenting a program which increments a counter for each line of code each time it is executed. Additionally, a program can be executed with input of varying size to highlight what lines show an exponential increase in executing as input size increases \cite{coppa2012input}. Profiler techniques generally require developers must frequently trace back through a program to understand what code is contributing to a bottleneck \cite{Jin12a}. 
Though code profiling can determine the location of a performance issue or bottleneck, it does not indicate what code change is required to improve performance. A performance improvement may not always be found at the same location. Finding a bottleneck does not always indicate the location of a solution to the bottleneck. This expresses the need for a technique which can determine more accurately where an improvement is likely to exist within a program.

As fault localisation \cite{jones2005empirical} has been used to automate bug fixing \cite{weimer2009automatically}, we seek similar methods for localising performance \cite{coppa2012input} to source code elements to benefit automated performance improvement \cite{langdon2013optimising,wu2015deep}. We seek localisation techniques which highlight code having the most effect on the overall execution cost of a program. Thus we are more interested in finding performance improvement opportunities than finding performance bottlenecks.

Our research question follows as:
\begin{itemize}
\item What performance localisation technique most accurately highlights locations of improvements?
\end{itemize}

Our hypothesis is that code locations which are particularly influential to program performance are likely good locations for finding performance improvements. 
To inspect this hypothesis, we consider to what extent code mutation can attribute performance to source code elements. Mutation has been used previously to find code locations which influence program performance \cite{wu2015deep} though this approach only considers mutations which leave program functionality unaffected. In this paper we look for mutation locations which reduce execution cost regardless of effect on functionality. As the goal is to use mutation to find hints for where a performance improvement may exist program, we exclude any mutations which reduce execution cost while leaving functionality ``correct''. 

We introduce 3 analysis techniques in \autoref{sec:perf-local-techn} based on two different types of mutation which highlight locations in code for their relevance to overall program performance. The first mutation approach deletes program statements (including any enclosed code) and measures the resulting savings in execution cost. The reduction in execution cost is attributed to deleted code. The second technique makes all possible changes to every modification location in a program and attributes performance change to these finer-grained modification points.

The intuition behind the use of mutation is that there are locations in a program which are ``levers on performance'' and have some disproportionately large control over execution cost. We are looking for code locations where small modifications produce a comparatively large change in overall program execution cost. These mutation-based approaches shift focus towards locations in code where modifications are likely to alleviate a bottleneck, performance bug or even some more modest improvement opportunity.

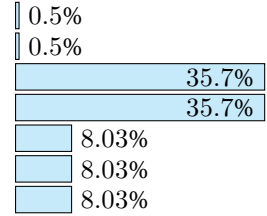
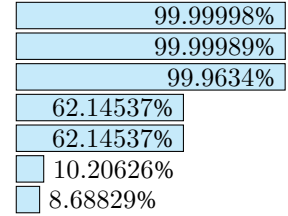
\begin{figure*}
  \begin{subfigure}[!t]{.45\linewidth}

      \centering
  \begin{lstlisting}[numbers=right]
void sort(Integer[] a,  int length){
 for (int h = 0; h < 2; h++) {
  for (int i=0; i < length; i++){
   for (int j=0; j < length - 1; j++){
    if (a[j] > a[j + 1]){
     int k=a[j];
     a[j]=a[j + 1];
     a[j + 1]=k;
}}}}}
\end{lstlisting}
  \caption{``BubbleLoops'' problem: Bubblesort with an extra redundant outer loop}
      \label{fig:bubbleloops}
\end{subfigure}\hspace{1.1cm}
  \begin{subfigure}[t]{.22\linewidth}
    \vspace*{-2.05cm}
    { 
    \begin{tikzpicture}[scale=1]
    \begin{axis}[ hide axis,
      xbar,
      width=5.5cm,
      height={ .02cm + ( 6.0 * .75cm ) },
      symbolic y coords={{l7},{l6},{l5},{l4},{l3},{l2},{l1}},
      xlabel={Pontérték \%},
      ytick=data, 
      nodes near coords={\pgfmathprintnumber[precision=5]\pgfplotspointmeta\%},
      visualization depends on={meta < 30 \as \valueissmall},
      every node near coord/.append style={
        anchor={\ifdim\valueissmall pt=1 pt west\else east\fi}
                },
      nodes near coords align = {horizontal}
      ]
      \addplot [draw=black, fill=cyan!20] coordinates {
        (8.03,{l7}) 
        (8.03,{l6}) 
        (8.03,{l5}) 
        (35.7,{l4})
        (35.7,{l3})
        (0.5,{l2}) 
        (0.5,{l1})  
      };
    \end{axis}
  \end{tikzpicture}
}
    \vspace{.25cm}
\caption{Profiler: Execution frequency for each statement}
    \label{fig:bubbleloopsprofile}
\end{subfigure}\hfill
  \begin{subfigure}[t]{.22\linewidth}
        \vspace*{-2.05cm}
{ \begin{tikzpicture}[scale=1]
    \begin{axis}[ hide axis,
      xbar,
      width=5.5cm,
      height={ .02cm + ( 6.0 * .75cm ) },
      symbolic y coords={{l7},{l6},{l5},{l4},{l3},{l2},{l1}},
      xlabel={Pontérték \%},
      ytick=data, 
      nodes near coords={\pgfmathprintnumber[precision=5]\pgfplotspointmeta\%},
      visualization depends on={meta < 30 \as \valueissmall},
      every node near coord/.append style={
        anchor={\ifdim\valueissmall pt=1 pt west\else east\fi}
                },
      ]
      \addplot [draw=black, fill=cyan!20] coordinates {
        (99.99998,{l1})
        (99.99989,{l2})
        (99.96340,{l3})
        (62.14537,{l4})
        (62.14537,{l5})
        (10.20626,{l6})
        (08.68829,{l7})
      };
    \end{axis}
  \end{tikzpicture}
}
\vspace{.25cm}
\caption{Deletion Analysis: Execution savings when a statement (including sub-statements) is deleted}
  \label{fig:bubbleloopsdeletion}
\end{subfigure}\hspace{.2cm}
\caption{BubbleLoops problem and profiles}
\label{bubbleExecutionProfile}
\small
\vspace{.35cm}
\end{figure*}

\noindent\textbf{Motivating example.}
The canonical example involves a variable which is initialised early in program execution and determines how many times a loop executes later in the program. The execution cost of this variable initialisation is low as the line is only executed once. However, a large amount of the overall execution cost of the program can be attributed to this initialisation if the variable is later used as a condition for how many times a loop executes.

To further illustrate the point we use a simple BubbleSort algorithm with a redundant outer loop added as seen in \autoref{fig:bubbleloops}. The additional outer loop does not change the program semantics but causes the BubbleSort algorithm to be needlessly iterated over a second time. A profiler will give this outer loop a very low value in terms of execution cost as can be seen in \autoref{fig:bubbleloopsprofile}, therefore taking attention away from a prominent performance improvement opportunity.

The execution count in \autoref{fig:bubbleloopsprofile} shows the number of times each statement is executed as a percentage. When an array of size 10 with all elements in reverse order is passed as \verb=a=, lines 4 \& 5 are both executed 200 times. The execution count for each statement will vary depending on the distribution of values within the array. An array of 10 values with a different ordering will produce a different execution count profile and can change the ranking of each statement with respect to the others. If a fully sorted array is passed then statements 5, 6 and 7 will not be executed. A reverse sorted array executes each line the maximum number of times possible and is expected to give the same ranking of statements as input size is increased. If profiling was used in this case to guide automated performance improvement \cite{langdon2013optimising}, it would appear to decrease the chances of finding this performance improvement as effort is spent modifying other locations.

In contrast, deletion analysis shows how much of the program execution cost is attributable to the outer loop.
\autoref{fig:bubbleloopsdeletion} shows the amount of execution cost that is saved when a statement (including any sub-statements) is removed. Execution cost savings are a percentage of the overall cost of executing the program. Note that as deletion analysis removes a statement inclusive of any sub-statements, percentages are cumulative. When statement 2 is removed the entire body of the method is removed, and so close to 100\% of the execution cost is saved. Statement 2 receives a marginally larger percentage and is ranked ahead of statement 3. Deleting line 6 will result in a program which does not compile, in this case the line receives the execution cost saving from its parent statement as all nodes within a statement (and any sub-statements) are given the same value initially.

\noindent\textbf{Contributions.}
In this paper we inspect the use of mutation to indicate the location of performance improvements in programs. We inspect two types of mutation:
\begin{itemize}
\item Deletion analysis takes advantage of the hierarchical nature of source code. Source code statements (and any child statements) are deleted and the resulting program variant is executed. 
\item Exhaustive mutation analysis makes all possible changes to each modification point in a program. For each code location, we can generate a set of program variants, one for each possible change at that location. In other words, we generate all possible first order mutants \cite{langdon:2015:csdc} for each modification point in a program. 
\end{itemize}

Using the results of these mutation techniques the characteristics of program variants are analysed for each location in the program. Using simple heuristics across the summarised information for each location, it is possible to determine the likelihood of a location being important for improving performance. We evaluate profiling against 3 different analysis approaches in \autoref{sec:perf-local-techn} for localising performance based on the previously mentioned mutation techniques:
\begin{itemize}
\item Under deletion analysis (\autoref{sec:deletion-analysis}), the difference in performance between the original and variant programs is attributed to all code which was deleted. Every statement in a program is tested in this way, giving code inside inner loops a lower value than their enclosing looping constructs . 
\item Under exhaustive mutation (\autoref{sec:exha-mutat-analys}), the number of times a program variant shows reduced execution cost is divided by the number of times a mutation results in a compiled program. 
\item Occasionally there is no possible single point mutation which will produce a compilable program. In this scenario we use the results of deletion analysis to fill in the gaps where exhaustive analysis was not able to glean any information (\autoref{sec:exha-delet-comb}). 
  \end{itemize}

We evaluate these approaches on a set of test problems and find:
\begin{itemize}
  \item profiling achieves the highest accuracy of all approaches on specific nodes, but does not generalise across our problem set (\autoref{tab:results})
  \item mutation analysis can, on average, better highlight locations of possible performance improvements in code (\autoref{fig:comparison})
  \item that there is a trade-off between the amount of computation required for each approach and the accuracy (\autoref{sec:comp-cost-analys})
\end{itemize}

\section{Performance Localisation Techniques Studied} 
\label{sec:perf-local-techn}
In this section we describe the localisation techniques that are compared in our evaluation.
Though many static analysis techniques exist for detecting performance issues, we compare with a profiling approach as recently used for automatic performance improvement \cite{langdon2013optimising}. 

\subsection{Profiling}
\label{sec:profiling}
Our approach to profiling is relatively fine-grained to other approaches which may measure, for example, elapsed time for method execution. We measure the number of times each statement in a program is executed. For each source code statement, as defined in the Java language specification \cite{javaspec}, we add an instrumentation statement as demonstrated in \autoref{fig:bubbleloopsinstrumented}. Each instrumentation statement consists of a function call with a program identifier and the line number for that location. When the instrumented program is run an execution count for each line is gathered.  

\begin{figure}[!b]
      \centering
      \begin{lstlisting}
void sort_5(Integer[] a,  int length){
 ASTInstrumenter.recordExecution(5,0)   
 for (int h = 0; h < 2; h++) {
  ASTInstrumenter.recordExecution(5,2)   
  for (int i=0; i < length; i++){
   ASTInstrumenter.recordExecution(5,3)   
   for (int j=0; j < length - 1; j++){
    ASTInstrumenter.recordExecution(5,4)   
    if (a[j] > a[j + 1]){
     ASTInstrumenter.recordExecution(5,5)
     int k=a[j];
     ASTInstrumenter.recordExecution(5,6)   
     a[j]=a[j + 1];
     ASTInstrumenter.recordExecution(5,7)   
     a[j + 1]=k;
    }
   }
  }
 }
 ASTInstrumenter.recordExecution(5,1)   
}
\end{lstlisting}
      \caption{Instrumented ``BubbleLoops'' problem, counting lines 0 to 7 for program variant ID 5}
          \label{fig:bubbleloopsinstrumented}
\end{figure}

\subsection{Deletion Analysis}
\label{sec:deletion-analysis}
Deletion analysis was designed in an attempt to shift focus from bottlenecks towards code which has some influence over performance. A program has a statement removed and the resulting program variant is evaluated. When a statement contains sub-statements, for example a ``FOR'', ``WHILE'' or ``IF'' statement, the inner block statement and all sub-statements are removed also. The hierarchical structure of imperative code is made accessible by using Abstract Syntax Tree (AST) parsers \cite{jdt} and is also used in scoping variables in code. Statements are removed in order of their appearance in a breadth-first approach as per AST structure. Outer loops are removed before inner loops, with the most nested code being removed last. 

Deletion analysis exploits the ordered and hierarchical structure of imperative code as execution cost is attributed to statements which appear earlier in the code and to statements higher in the hierarchy. For example, the body of a "FOR" loop which may contain many statements can be considered the child of a "FOR" statement.
The parent "FOR" statement is attributed the execution cost of all child statements, however deeply nested, by deleting the "FOR" statement including all child statements and measuring the execution cost reduction of the variant program when compared with the original. 

Deletion analysis may not always be applicable for every statement in a program. Consider statement 6 in \autoref{fig:bubbleloops} which initialises the variable \verb=k=. Deleting this line will result in a program which does not compile and which we cannot evaluate. Our choice in this scenario is to either attribute zero to this statement or attribute a ranking by way of its enclosing statement higher in the hierarchy (in this case it would receive a value of 62.15 from statement 5). 

\subsection{Exhaustive Mutation Analysis}
\label{sec:exha-mutat-analys}
Deletion analysis is not always able to directly attribute information about the cost associated with all statements in a program. The approach also gives all code elements within a statement the same value. We are thus motivated to inspect mutation of code elements at a finer level than statement. This approach will give different rankings to individual code elements within a statement and for statements which could not be evaluated under deletion.
It is specifically designed to perform a type of performance sensitivity analysis by way of mutation. For example, any valid change to variable initialisation or the loop condition in the outermost loop (statement 2 in \autoref{fig:bubbleloops}) is likely to show a pronounced change in the execution cost of the resulting program. 

A set of program variants is produced by repeatedly replacing a node with every other valid alternative node as found in the original program.
A program P, is made up of a set s of all elements in the program. Let s' be a set of clones of all elements in s. For each element l in s in the program P, a variant program can be generated by exchanging l for each of the elements in s'. In other words, exhaustive mutation analysis takes all code elements in a program and replaces them with all alternatives. Alternative code elements are gleaned from the program itself but we add all language-defined \cite{javaspec} operators regardless of whether they are contained in the program.  While this appears to produce a large number of costly program evaluations, in practice not all variant programs are compilable or evaluatable as shall be seen in \autoref{sec:comp-cost-analys}.

Each node in a program is attributed a value by taking the number of times a modification resulted in a program with reduced execution cost divided by the number of times a modification resulted in a compilable variant as written in \autoref{eq:nodeval}.

Example results of values attributed by exhaustive mutation analysis to statement 2 in \autoref{fig:bubbleloops} \verb$for (int h = 0; h < 2; h++) {$  are shown in \autoref{tab:exhaustiveexample}. It is not possible to modify some nodes in the AST as listed in the table by ``-''. Where no mutation can be produce a compilable program variant, the value is 0.

\begin{table}[!h]
  \begin{center}
    \caption{Exhaustive mutation analysis example on a single line of code taken from the BubbleLoops problem}
    \begin{tabular}{|l|l|c|}
      \hline
      \textbf{Node Num} & \textbf{Textual representation} & \textbf{Value} \\
      \hline
      1 & \verb$for (int h...$  & 1.0 \\
      2 & \verb$int h=0$ & -\\
      3 & \verb$h=0$ & - \\
      4 & \verb$h$ & .6\\
      5 & \verb$0$ & .7\\
      6 & \verb$h < 2$ & - \\
      7 & \verb$h$ & .16\\
      8 & \verb$2$ & .85\\
      9 & \verb$h++$ & - \\
      10 &\verb$h$ & 0 \\
      \hline
    \end{tabular}
    \label{tab:exhaustiveexample}
  \end{center}
\end{table}

\begin{eqcap}
\begin{equation}
  NodeVal = \frac{N_{execution reduction}}{N_{compiled}}
  \label{eq:nodeval}
\end{equation}
\caption{Exhaustive Analysis gives each node a quotient value of the number of times the execution cost is reduced, divided by the number of times a compilable program is created.}
\end{eqcap}

\subsection{Exhaustive and Deletion Combined}
\label{sec:exha-delet-comb}
Even though exhaustive mutation analysis makes program modifications at a sub-statement level, it is still possible that no single element modification is able to produce a compilable, and hence evaluatable, program variant. Where no compilable program can be produced by modifying a particular location in a program we are missing information about it's relevance. This can occur where variable scoping prevents replacement by any other variable. To alleviate this issue, we use the results of deletion analysis to ``fill the gaps'' in the results of exhaustive mutation analysis where no single change produced a compilable program.

\section{Methodology}
To evaluate the analysis techniques, we use a set of problems with known performance improvements. By comparing programs with their variants which contain known performance improvements we can find what code elements differ. The nodes that differ between programs are considered improvement opportunities in the inefficient versions of a program. These ``improvement'' nodes are of highest importance and receive the highest rank among all nodes in the program. The highest ranked and therefore most important nodes are those which are required to change to improve the performance of a program. We apply performance localisation techniques to these problems and gather node rankings. We compare the node rankings to our idealised ranking to determine which mutation technique is most accurate.

\subsection{Program Evaluation Measures}
We measure performance in terms of execution cost. For profiling, this is the number of statements executed when the program is run. For mutation approaches, we use a more fine-grained measure of execution cost by counting the number of byte-codes executed by the JVM \cite{kuperberg2008a}.

\subsection{Problem Set}
\label{sec-2}
We use a variety of sort implementations and a Huffman code-book (or ``dictionary'') generation implementation to observe how profiling and mutation analysis can find improvement locations. A long-form code listing of all programs in our problem set is available \footnote{\url{https://www.scss.tcd.ie/~codykenb/locoGP-ImprovementsFound.html}}.

\begin{table*}[!ht]
  \begin{center}
    \caption{Problem Improvement Overview}
    \begin{tabular}{|l|c|c|c|c|c|}
      \hline
      \textbf{Problem Name}  & \textbf{LOC} & \textbf{AST Nodes} & \textbf{Imp Nodes}& \textbf{Imp}& \textbf{Improvement Type}\\
      \hline
      Insertion Sort & 13& 60& 3 & 9\% & Loop unrolling \\ 
      \hline
      Bubblesort & 13 & 62 & 5  & 45\% & Redundant Traversal (exclude sorted portion)\\ 
      \hline
      BubbleLoops \tnote{This is an artificial performance bug, Bubblesort with an additional outer redundant loop} & 14 & 72 & 8  & 71\% & Redundant Traversal (exclude sorted portion) \\ 
      \hline
      Selection Sort 2 & 16 & 72& 1 & 11\% & Removed redundant increments during tests \\ 
      \hline
      Selection Sort & 18 & 73& 1& 2\% &  Removed redundant array access\\ 
      \hline
      Shell Sort & 23 & 85& 3 & 5\% & Various changes in increment size\\ 
      \hline
      Radix Sort & 23 & 100 & 3& 3\% & Reduced iteration, comparison with 0 \\ 
      \hline
      Quick Sort & 31 & 116 & 2  & 54\% & Reduced iterations, remove tests\\ 
      \hline
      Cocktail Sort & 30 & 126 & 2& 15\% & Cloned and perforated loops (loop unrolling)\\ 
      &  &  & &  & Redundant Traversal (exclude sorted portion)\\ 
      \hline
      Merge Sort & 51 & 216& 1& 5\% & Remove redundant array clone\\ 
      \hline
      Heap Sort &62 & 246 &2  & 41\% & Remove redundant array access and assignment\\ 
      \hline
      Huffman Code-book & 115 & 411 & 5  & 43\% & Same as Bubblesort \\ 
      \hline
    \end{tabular}
    \label{table:GPimprovements}
  \end{center}
\end{table*}

\autoref{table:GPimprovements} lists the implementations used and provides descriptive measures for each program as well as improvement types:
\begin{itemize}
\item \textbf{LOC} refers to the number of lines of code in the program
\item \textbf{AST nodes} refers to the number of modification points in each program when it is parsed into an Abstract Syntax Tree representation \cite{jdt}.
\item \textbf{Imp Nodes} refers to the number of nodes or locations in the program which need to be changed to achieve an improved version of the program. Although there are a number of different code modifications which can yield an improved variant of a program, we use the improvements which give the greatest reduction in execution cost with the smallest number of modifications. Modifications which reduce program functionality the least when applied individual are also favoured. We thus use the improvements that are the 'easiest' or most probable set of modifications to be found with a search algorithm. This includes nodes involved in multiple known improvements.
\item \textbf{Improvement} refers to the largest percentage improvement in execution cost known for each program \cite{Cody-Kenny:2015:gi}.
\item \textbf{Improvement Types} gives a high level description of the known improvement types for each program. For loop unrolling, the important node is the containing block statement. 
\end{itemize}

\subsection{Test Cases}
The execution cost of the test problem implementations we consider are affected by input size and distribution. We use a range of input sizes and distributions to ensure the profile is general. Input array size for sort implementations ranges from one to ten values. The distribution includes random, fully and reverse sorted ordering. For the Huffman code-book problem five different test cases which include arrays with repeated sequences and those without any repeated character.

\subsection{Comparing Localisation Techniques}
\label{sec:comp-local-techn}
We have an idealised ``best'' ranking of nodes which put nodes involved in some improvement at the top which we term ``improvement'' nodes. These top ranked improvement nodes are required to change to produce a known improvement in each program.
Fractional ranking is used as all nodes in a statement jointly share a given ranking. E.g. if two nodes share first place, then both nodes are given the ranking of "1.5" as this will be the ranking of the nodes on average if they are selected randomly.

For each program, each localisation technique produces a ranking for all nodes. Only a small proportion of these nodes are required to change to improve the performance of a program. The closer an improvement node is to where it should be in the idealised ranking is used as our measure of ``accuracy''. The distance an improvement node is from where it should be in our idealised ranking is what we use as our ``ranking error'' measure. We normalise the ranking error for each node by dividing it by the number of nodes in the overall program to find at what percentile the node is placed. 

For each technique we compare the percentile ranking error of each important node across all problems. This gives us 48 important nodes across all problems for comparison. We also do pair-wise comparison between techniques to be sure there is a statistically significant difference between them directly. We find the different between the approaches by subtracting the percentile rankings of the important nodes. We use a bootstrapping technique to analyse these differences. We sample randomly from these differences 100 times, with replacement, and calculate the average. This is repeated 100 times. This bootstrap approach gives an estimator for mean and approximate 95\% confidence interval are given by the 0.015 and 0.975 quantiles.  

We further summarise results be looking at nodes ranked in the upper 50 percentile of all the nodes represent instances where profiling has accurately highlighted the location of an improvement. We chose the 50 percentile as a simple way to show how the nodes in a program can be segregated. How the ranking of nodes is used to infer importance is dependent on the probability distribution generated from this ranking. The 50 percentile represents the median of node rankings with nodes in the upper half being considered more important than those in the lower half. Nodes which have a ranking in the upper 50 percentile of all nodes represent instances where profiling can be said to have been ``accurate''. As increasing the ranking of one node reduces the ranking of another, where an improvement node is in the lower 50 percentile of all nodes then the technique can be said to be ``deceived''.

A normalised percentile ranking error is the distance a node is ranked from its ideal ranking, divided by the number of nodes in that program (\autoref{eq:functionality}).
\begin{eqcap}
\begin{equation}
PercentRankError = \frac{  R_{i}-R_{l}} {N_{total}}
\label{eq:functionality}
\end{equation}
\caption{Percentile Ranking Error measure calculated for each improvement node.}
\end{eqcap}

\section{Results}

The four performance localisation techniques, Profiling, Deletion, Exhaustive and Deletion with Exhaustive gap filling (Ex \& De), are compared for accuracy in \autoref{tab:results}. 

We show a split at the 50 percentile to make the point that using a probability distribution over these accuracy values will result in some cutoff point where nodes below will receive lower importance and those above will receive higher importance (in comparison to a scenario where all nodes have the same ranking or importance). We can conceive of importance being only those nodes which are in the top 1\% of all nodes. In such a scenario, profiling is the only approach which would designate any node as important. Profiling would be considered best in this scenario but would only highlight a single improvement node as important. The lower we place the threshold for importance as a percentile, the larger the combinations of those nodes become. The more of the important nodes we want to include as important, the more program nodes we must consider. To include all important nodes we must consider all nodes in the program, which does not help us reduce the number of nodes worth considering important. The more nodes we consider, the exponentially more combinations we need to consider. 

When interpreting \autoref{tab:results} we consider Exhaustive with Deletion (Ex \& De) to be the best as this approach places the largest number of improvement nodes in the upper 50 percentile. The three mutation-based approaches also put a majority of the improvement nodes in the upper half of all nodes. 

\begin{table}[!h]
\caption{\label{tab:results} The accuracy of performance localisation techniques. }
\begin{tabular}{|r|c|c|c|c|}
\hline
Accuracy & Profiler & Deletion & Exhaustive & Ex \& De\\
\hline
99-100\% &        1  & 0   &   0 & 0  \\
90-99\%  &        7  & 8   &   9 & 11  \\
80-90\%  &        7  & 2   &   9 & 6  \\
70-80\%  &        3  & 10  &   7 & 5  \\
60-70\%  &        3  & 6   &   5 & 9  \\
50-60\%  &        2  & 4   &   2 & 5  \\
\hline                            
40-50\%  &        2  & 3   &   4 & 1  \\
30-40\%  &        6  & 2   &   5 & 3  \\
20-30\%  &        10 & 5   &   1 & 6  \\
10-20\%  &        2  & 2   &   0 & 2  \\
0-10\%   &        5  & 6   &   6 & 0  \\
\hline
\end{tabular}
\end{table}

We further show a pair-wise comparison of the approaches using a bootstrap statistical technique (as described in \autoref{sec:comp-local-techn}) over the differences of percentiles for each improvement node. \autoref{fig:comparison} shows the difference between Profiling and Deletion, Deletion and finally Exhaustive and Exhaustive with Deletion. On average, improvement nodes are ranked roughly 2.75 percentage points higher under Deletion analsis when compared with a Profiler, 6.25 percentage points higher under Exhaustive analysis when compared with Deletion, and 3.6 percentage points higher still when using Exhaustive with Deletion. 

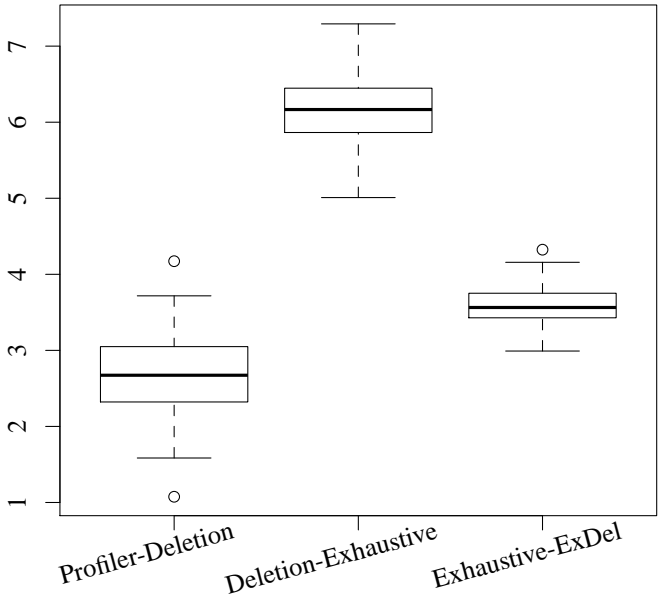
\autoref{fig:comparison} also cross validates our evaluation as the differences in improvement node percentiles correlate with the ordering (though not magnitude) of which techniques are more accurate than others in \autoref{tab:results}. The difference between the number of improvement nodes ranked in the upper half of all nodes as shown in \autoref{tab:results} (Deletion ranks more nodes in upper half than Profiling, Exhaustive more than Deletion, and Exhaustive \& deletion gap filling more than Exhaustive alone).   

\begin{figure}[!h] 
  \caption{Comparison of the \textbf{differences} between node percentile rankings for the four different approaches}
  \tikzset{every picture/.style={scale=0.9}}
  \hspace{-1.1cm}
  \noindent
\begin{tikzpicture}[x=1pt,y=1pt]
\definecolor{fillColor}{RGB}{255,255,255}
\path[use as bounding box,fill=fillColor,fill opacity=0.00] (0,0) rectangle (325.21,325.21);
\begin{scope}
\path[clip] ( 49.20, 61.20) rectangle (300.01,276.01);
\definecolor{drawColor}{RGB}{0,0,0}

\path[draw=drawColor,line width= 1.2pt,line join=round] ( 66.23,120.27) -- (128.16,120.27);

\path[draw=drawColor,line width= 0.4pt,dash pattern=on 4pt off 4pt ,line join=round,line cap=round] ( 97.20, 85.46) -- ( 97.20,109.03);

\path[draw=drawColor,line width= 0.4pt,dash pattern=on 4pt off 4pt ,line join=round,line cap=round] ( 97.20,153.67) -- ( 97.20,132.31);

\path[draw=drawColor,line width= 0.4pt,line join=round,line cap=round] ( 81.71, 85.46) -- (112.68, 85.46);

\path[draw=drawColor,line width= 0.4pt,line join=round,line cap=round] ( 81.71,153.67) -- (112.68,153.67);

\path[draw=drawColor,line width= 0.4pt,line join=round,line cap=round] ( 66.23,109.03) --
	(128.16,109.03) --
	(128.16,132.31) --
	( 66.23,132.31) --
	( 66.23,109.03);

\path[draw=drawColor,line width= 0.4pt,line join=round,line cap=round] ( 97.20,168.24) circle (  2.25);

\path[draw=drawColor,line width= 0.4pt,line join=round,line cap=round] ( 97.20, 69.16) circle (  2.25);

\path[draw=drawColor,line width= 1.2pt,line join=round] (143.64,232.03) -- (205.57,232.03);

\path[draw=drawColor,line width= 0.4pt,dash pattern=on 4pt off 4pt ,line join=round,line cap=round] (174.61,195.00) -- (174.61,222.37);

\path[draw=drawColor,line width= 0.4pt,dash pattern=on 4pt off 4pt ,line join=round,line cap=round] (174.61,268.06) -- (174.61,241.01);

\path[draw=drawColor,line width= 0.4pt,line join=round,line cap=round] (159.13,195.00) -- (190.09,195.00);

\path[draw=drawColor,line width= 0.4pt,line join=round,line cap=round] (159.13,268.06) -- (190.09,268.06);

\path[draw=drawColor,line width= 0.4pt,line join=round,line cap=round] (143.64,222.37) --
	(205.57,222.37) --
	(205.57,241.01) --
	(143.64,241.01) --
	(143.64,222.37);

\path[draw=drawColor,line width= 1.2pt,line join=round] (221.05,148.77) -- (282.98,148.77);

\path[draw=drawColor,line width= 0.4pt,dash pattern=on 4pt off 4pt ,line join=round,line cap=round] (252.02,130.45) -- (252.02,144.43);

\path[draw=drawColor,line width= 0.4pt,dash pattern=on 4pt off 4pt ,line join=round,line cap=round] (252.02,167.77) -- (252.02,154.76);

\path[draw=drawColor,line width= 0.4pt,line join=round,line cap=round] (236.54,130.45) -- (267.50,130.45);

\path[draw=drawColor,line width= 0.4pt,line join=round,line cap=round] (236.54,167.77) -- (267.50,167.77);

\path[draw=drawColor,line width= 0.4pt,line join=round,line cap=round] (221.05,144.43) --
	(282.98,144.43) --
	(282.98,154.76) --
	(221.05,154.76) --
	(221.05,144.43);

\path[draw=drawColor,line width= 0.4pt,line join=round,line cap=round] (252.02,173.08) circle (  2.25);
\end{scope}
\begin{scope}
\path[clip] (  0.00,  0.00) rectangle (325.21,325.21);
\definecolor{drawColor}{RGB}{0,0,0}

\path[draw=drawColor,line width= 0.4pt,line join=round,line cap=round] ( 97.20, 61.20) -- (252.02, 61.20);

\path[draw=drawColor,line width= 0.4pt,line join=round,line cap=round] ( 97.20, 61.20) -- ( 97.20, 55.20);

\path[draw=drawColor,line width= 0.4pt,line join=round,line cap=round] (174.61, 61.20) -- (174.61, 55.20);

\path[draw=drawColor,line width= 0.4pt,line join=round,line cap=round] (252.02, 61.20) -- (252.02, 55.20);

\path[draw=drawColor,line width= 0.4pt,line join=round,line cap=round] ( 49.20, 66.75) -- ( 49.20,258.70);

\path[draw=drawColor,line width= 0.4pt,line join=round,line cap=round] ( 49.20, 66.75) -- ( 43.20, 66.75);

\path[draw=drawColor,line width= 0.4pt,line join=round,line cap=round] ( 49.20, 98.74) -- ( 43.20, 98.74);

\path[draw=drawColor,line width= 0.4pt,line join=round,line cap=round] ( 49.20,130.73) -- ( 43.20,130.73);

\path[draw=drawColor,line width= 0.4pt,line join=round,line cap=round] ( 49.20,162.72) -- ( 43.20,162.72);

\path[draw=drawColor,line width= 0.4pt,line join=round,line cap=round] ( 49.20,194.71) -- ( 43.20,194.71);

\path[draw=drawColor,line width= 0.4pt,line join=round,line cap=round] ( 49.20,226.70) -- ( 43.20,226.70);

\path[draw=drawColor,line width= 0.4pt,line join=round,line cap=round] ( 49.20,258.70) -- ( 43.20,258.70);

\node[text=drawColor,rotate= 90.00,anchor=base,inner sep=0pt, outer sep=0pt, scale=  1.00] at ( 34.80, 66.75) {1};

\node[text=drawColor,rotate= 90.00,anchor=base,inner sep=0pt, outer sep=0pt, scale=  1.00] at ( 34.80, 98.74) {2};

\node[text=drawColor,rotate= 90.00,anchor=base,inner sep=0pt, outer sep=0pt, scale=  1.00] at ( 34.80,130.73) {3};

\node[text=drawColor,rotate= 90.00,anchor=base,inner sep=0pt, outer sep=0pt, scale=  1.00] at ( 34.80,162.72) {4};

\node[text=drawColor,rotate= 90.00,anchor=base,inner sep=0pt, outer sep=0pt, scale=  1.00] at ( 34.80,194.71) {5};

\node[text=drawColor,rotate= 90.00,anchor=base,inner sep=0pt, outer sep=0pt, scale=  1.00] at ( 34.80,226.70) {6};

\node[text=drawColor,rotate= 90.00,anchor=base,inner sep=0pt, outer sep=0pt, scale=  1.00] at ( 34.80,258.70) {7};

\path[draw=drawColor,line width= 0.4pt,line join=round,line cap=round] ( 49.20, 61.20) --
	(300.01, 61.20) --
	(300.01,276.01) --
	( 49.20,276.01) --
	( 49.20, 61.20);
\end{scope}
\begin{scope}
\path[clip] (  0.00,  0.00) rectangle (325.21,325.21);
\definecolor{drawColor}{RGB}{0,0,0}

\node[text=drawColor,rotate= 15.00,anchor=base east,inner sep=0pt, outer sep=0pt, scale=  1.00] at (122.75, 51.74) {Profiler-Deletion};

\node[text=drawColor,rotate= 15.00,anchor=base east,inner sep=0pt, outer sep=0pt, scale=  1.00] at (207.91, 51.74) {Deletion-Exhaustive};

\node[text=drawColor,rotate= 15.00,anchor=base east,inner sep=0pt, outer sep=0pt, scale=  1.00] at (285.32, 51.74) {Exhaustive-ExDel};
\end{scope}
\end{tikzpicture} 
  \vspace{-1cm}
  \label{fig:comparison}
\end{figure}

\begin{table*}[!ht]
    \begin{center}
\caption{\label{tab:summary}Summary}
\begin{tabular}{|l|l|l|l|l|}
\hline
 & Profiler & Deletion & Exhaustive & Ex \& Del\\
\hline
Nodes most accurate (out of 48) & \textbf{13} & 10 & \textbf{13} & 12\\
Nodes least accurate & 15 & 19 & 8 & \textbf{6} \\
Nodes ranked in upper half & 23 & 30 & 32 & \textbf{36}\\
Nodes ranked in lower half & 25 & 18 & 16 & \textbf{12}\\
Problems with only accurate nodes (out of 12) & 4 & 4 & \textbf{5} & \textbf{5}\\
Problems majority nodes deceived & 6 & \textbf{2} & \textbf{2} & \textbf{2}\\
Best on Problems & 3 & 1 & \textbf{4} & \textbf{4}\\
\hline
\end{tabular}
  \end{center}
\end{table*}

Table \ref{tab:summary} shows descriptive summary values for each technique (Long form results are available \cite{DBLP:journals/corr/Cody-KennyB16}. \\ \textbf{Nodes most accurate} shows for how many nodes each technique is the most accurate of all techniques. Profiling has the highest accuracy values on the most nodes (13 out of a total of 48 important nodes). \\
\textbf{Nodes least accurate} sums the number of improvement nodes each technique attributes the lowest ranking of all techniques. Deletion analysis gives the lowest ranking to the most nodes when compared with all other techniques. \\
\textbf{Nodes ranked in upper half} and \textbf{Nodes ranked in lower half} show a sum of the number of nodes ranked above and below the 50 percentile respectively. \\
\textbf{Problems with only accurate nodes} counts the number of problems which do not contain any improvement nodes ranked in the lower 50 percentile.\\
\textbf{Problems majority nodes deceived} shows a less stringent count of the number of problems where a majority of the improvement nodes are ranked in the lower half of all nodes. Where a majority of nodes are given a low ranking a technique can be said to be ``deceived'' as to the location of an improvement. \\
\textbf{Best on Problems} refers to when a technique has gives a majority of nodes the highest ranking.

Although profiling is accurate on 23 nodes across 8 problems, it is also deceived on a majority of the improvement nodes for 6 of the problems. It did however perform better than any other approach on 3 of the 12 problems including the Huffman Code-book problem which is the largest in our test set.

The use of Exhaustive analysis with deletion refinement ("Ex \& Del" in table \ref{tab:summary}) was least deceived of all techniques across all nodes, with only 12 nodes lower than the 50 percentile of all nodes. It was deceived on at least 1 node in 7 of the 12 problems and was deceived on the majority of important nodes in 2 of the problems. It also has the highest accuracy on 12 of the 48 important nodes. It performs the best across 4 of the problems.

In these results we assume that if an approach is deceived on a majority of important nodes in a problem, it is likely that it will take longer to find an improvement as GP modifies other locations in the program. If half or more of the nodes are ranked highly, then it is likely that the approach will help GP find at least one of the possible improvements more quickly. 

We consider a technique's ability to avoid being deceived as being more important than being the most accurate. We expect that there is some threshold value below which the use of a technique to guide a search process would lower the chances of finding an improvement. This threshold value is likely to have a polarising effect on a search algorithm. Effort spent modifying irrelevant nodes is effort that is not spent on important nodes. Due to this, we can hypothesise that a search algorithm would be delayed in finding performance improvements when focusing too much search effort on irrelevant nodes.

This is most obviously exemplified in the hand-crafted "BubbleLoops" problem, where an extra redundant outer loop has been added to Bubblesort. Profiling attributes a very low ranking to locations where simple changes which would half the execution cost of the program. Other examples which were not specifically crafted to be deceptive problems include Selection 2, Selection, Shell, Radix and Cocktail sort.

\subsection{Computational Cost of Analysis}
\label{sec:comp-cost-analys}
Profiling is the cheapest analysis to perform as instrumentation only need be performed once and a single execution of a program is needed to gather results. Even if we use profiling to find each statement's sensitivity to input size \cite{coppa2012input}, we may only need use a small number of test cases to find this information. As evaluation time dominates, we use this as our measure of computational cost. We take profiling to cost a single evaluation. 

When a program is mutated, the possible evaluation categories a variant program may fall into:
\begin{enumerate}
    \item  Not Compilable
    \item  Infinite Loop
      \item  Run-time Error where functionality \& execution cost differs from original program
    \item  Functionally degraded where functionality \& execution cost differ
    \item  More expensive (execution cost only differs from original program)
    \item  Identical to original program in terms of functionality and execution cost measures
    \item  Less expensive in terms of execution cost
\end{enumerate}

Previous results indicate that a large proportion (71 - 84\%) of variant Java programs do not compile \cite{Cody-Kenny:2015:gi} although these values are found under a wider range of mutations than considered here (statement cloning is allowed). 

All statements in a program can be legally deleted per the Java syntax due to the context insensitive nature of it's grammar regarding statements. Deletion analysis, as we implemented it, requires instrumentation for every variant program. As we create a variant program by deleting each statement the number of evaluations required is almost linear to the number of statements in a program. In practice it is slightly less than linear as deleting some lines of code results in a program variant which does not compile and does not need to be evaluated. Evaluating whether a program does not compile is quicker relative to the time it takes to fully evaluate a runnable program. 
Deletion based localisation strikes a balance between being relatively accurate across many problems and having an execution cost linear with program size.

On the face of it, exhaustive analysis for a program containing n elements gives n\\! combinations. Evaluation is not required where a single point mutation is not possible due to the Java type system. 
The existince of duplicate code elements in a program reduces the number of variants that need to be evaluated also. Less still, are the number of programs which compile. Unfortunately we can not exclude programs with infinite loops\footnote{As we cannot determine for how long a program will execute, we somewhat arbitrarily choose a practical timeout of 2.5 times the program's execution.} and runtime errors \footnote{We could conceivably estimate the characteristics of a program which shows a run-time error on certain input values but runs successfully on other input values, for example, smaller input values or values which are already sorted which cause less of the code to be executed are less likely to cause run-time issues. This information is telling in itself, and could also represent an additional dimension to location analysis.}. In any case, exhaustively mutating all code elements with all other elements in a program is practical for the relatively small programs in our test set. Although many replacements are not possible due to language typing constraints as enforced by the AST representation used, exhaustive mutation remains expensive, requiring the attempted replacement of every node with every other node. Many replacements will result in programs which can be quickly found to not compile, and therefore do not incur the comparatively large evaluation cost of repeat variant program execution with several different test input values.

\begin{figure}[!h] 
  \caption{Comparison of the cost of analysis between Exhaustive, Deletion and Profiling. Profiling is flat, requiring only one evaluation. Deletion is linear with program size. Exhaustive is exponential in relation to the number of AST nodes free to be modified. }
  \tikzset{every picture/.style={scale=0.84}}
  \hspace{-.2cm}
  \noindent
  \input{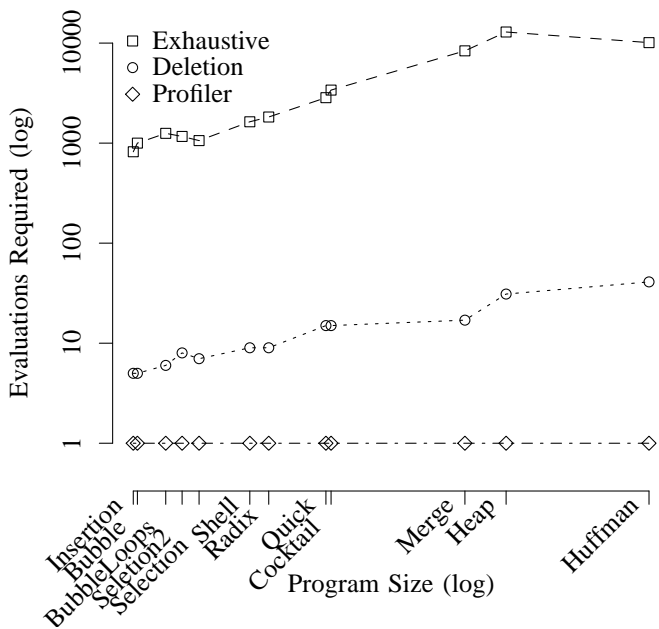} 
  \label{fig:analysiscost}
\end{figure}

\subsection{Threats to validity}
\label{sec-9}

The main threat to validity of our results is the size of the problem set with the concern being that our results do not generalise outside this set. This issue is of particular concern due to the limited variety of program type in our set; all but one of our test programs implements a sorting algorithm. Though the problem set of Sort and Huffman Codebook problems appears to be varied enough to make ranking improvement nodes highly across all problems currently unattainable, there remains a potential issue that the approach of exhaustive mutation and deletion analysis has been specialised to the algorithms in our problem set. Adding problems to the test set with particular attention paid to choosing a wider variety of problem types would reduce this concern. The length of programs is relatively small which calls into question how accuracy is affected when analysis is performed over much larger programs. As we use a sum total of execution cost it may be more difficult to measure how a mutation affects the overall cost.

The important nodes listed in our tables are sometimes part of multiple possible improvements. There are dependencies amongst some of the nodes where modifications must be made in a certain sequence to yield an improved program making some improvements easier to find than others. Not all nodes are equal, given that a change in some may produce low functionality programs and are dependent on other modifications. As not all nodes are equal in terms of dependencies a simple summation summary may not appropriately capture a localisation techniques accuracy. If a majority of important nodes in a program are highly accurately identified it may not improve search where these nodes depend on one specific node which has unfortunately been misidentified. The "importance" of nodes is thus not uniform. This concern can be addressed by a closer inspection of how ``difficult'' each improvement is to achieve. If an improvement requires multiple changes to the program it can be said to be more difficult to find than an improvement requiring only a single modification. 

\section{Discussion}

The major advantage of Profiling is the relatively low computational cost required. A single run of an instrumented program is enough to profile. 
Deletion analysis requires a program execution for each statement in a program though is less deceived on average than a profiler. Exhaustive mutation is more accurate still but also considerably more expensive to perform. The major disadvantage of using mutation is the high computational cost to perform localisation. 

The problem size we use is relatively small and there is a potential limitation especially with exhaustive mutation regarding scalability. A potential solution might be to use a hybrid of approaches. Deletion analysis could be used initially to find what statements influence execution cost the most. We could only perform deletion analysis for code blocks which appear to be worth it. If removing the outermost loop reduces execution cost by some small fraction of overall cost, it may not be worth deleting and executing further nested statements. At some depth in the program subtree deletion analysis can be skipped where execution cost savings are negligible. Once deletion analysis has identified the most influential lines of code, exhaustive mutation can then be used sparingly to only distinguish between nodes within highly influential statements. Such an approach would further exploit the hierarchical structure of source code. 

We use our results to say that the location of a performance bottleneck, as typically found using a profiler, does not always highlight potential performance improvements. When a performance improvement receives a low ranking, a search algorithm such as genetic programming will be less likely to find the improvement than had there been no node ranking at all.

The cost of performing mutation can be offset in scenarios when mutation is performed for other purposes such as mutation testing \cite{jia2011analysis} or genetic improvement \cite{weimer2009automatically,petke2014using}. Our results in this paper show that it is worth further attempting to further exploit the information generated by repeatedly executing mutated programs. Our main use case for this approach is as a guide for Genetic Programming (GP) to find performance improvements \cite{Cody-Kenny:2015:gi}. As mutation is the main driving force of the GP search process, performance localisation can potentially guide GP to performance improvements more quickly.

\section{Related Work}
\label{sec-10}
If we consider multiple versions of large programs as program variants then we can say that many modified versions of a program have been used to find performance issues on large scale software \cite{nagaraj2012structured}.
Using mutation to create many program variants has long been studied for software testing \cite{jia2011analysis} and has recently been inspected for understanding the robustness of software \cite{schulte2012software,langdon:2015:csdc}.

Closer to our approach is the use of mutation to discover ``deep parameters'' or locations where a modification in code relevant to program performance \cite{wu2015deep}. A deep parameter is a program mutation which affects program performance but not functionality. If program functionality changes in any measurable way, per an available test suite, the code location which was modified is removed from consideration as an interesting location for performance improvement. In contrast, our work shows that there is value in considering the location of mutations which degrade functionality but crucially also reduce execution cost. 

The implicit nature of performance means it can be difficult to understand the nuances and interactions between program source code, input values and execution environment. This is especially true in many JavaScript environments where execution of the same code can differ widely \cite{selakovic2016performance}. A code change can provide better performance in some environments but can reduce performance in others. A reasonable amount of improvements actually reduce performance \cite{selakovic2016performance}. This may be due to a perceived opportunity that is not actually validated with any empirical evidence or an improvement that is beneficial in one environment but increases execution cost in another. This makes the case for using more automated tooling to localise performance and aid program understanding.

Many traits of a program (or ``program spectra'') can be measured to decompose a program internally \cite{harrold1998empirical}. 
A prominant approach to aid program understanding is to count the execution frequency of source code objects which is commonly referred to as profiling \cite{ball1999concept}. Profiling is widely used to find performance ``bottlenecks'' in code  and highlights where the symptoms of performance issues can be observed \cite{conf/issta/ShenLPG15,Jin12a,conf/icsm/VasquezVLP15,conf/icse/ManotasPC14}. Many works build on the basic concept of profiling by using a range of input values to decompose and seperate the parts of a program and has been shown to be scalable to large numbers of inter-dependent input values \cite{grechanik2012automatically}. 
Input sensitive profiling uses progressively larger sized input values to highlight what lines of code have a particularly acute response to increased program input size \cite{coppa2012input}. A similar approach has been used with success to guide GP \cite{langdon2013optimising} showing the importance of performance localisation. 

Randomised input values can produce different executation traces which are used to isolate the root-cause of a performance issue. By finding a baseline expected performance on some input, subsequent input values can expose instances where performance is outside baseline. Root-cause can be determined by correlating system state against these anomalous instances \cite{killian2010finding}. Root-cause is pinpoonted by finding the ``divergent point'' where an anomalous execution deviates from expected. From this point to the point where a performance bug manifests can be considered suspicious. This approach was tested on large scale distributed systems.

Although there is no clear definition of what constitutes a performance bug we can generally say that a bug of this type is encountered when a program takes prohibitively longer than is expected or necessary to produce an output \cite{nistor2013discovering}. Some definitions use the existence of quadratic or cubic asymptotic execution response as input size increases as a clear indicator of a performance bug. Thus we feel that the definition of what constitutes a performance bug to be somewhat subjective. As we would like to have all programs execute as quickly as possible and the absence of a performance bug should not prevent us from attempting to improve program performance. A useful trait of profiling for improvement not specific to performance bug finding is that it provides a ranking of all code elements in a program as they contribute to program execution cost. A full ranking of code gives an ordering to what statements contribute the most without imposing a strict cutoff for what constitutes a ``performance bug''.  Conversely, where a quadratic or cubic execcution response may be optimal for a specific algorithm and input, it would be unfair to classify problems which there are no known sub-exponential solutions \cite{impagliazzo1998problems} as performance bugs due to programmer error \cite{nistor2013discovering}. 

Static analysis is a lightweight alternative to dynamic analysis for finding performance issues. Static analysis appears to be more specific to certain types of performance issues\cite{conf/pldi/OlivoDL15,nistor2015c}.   
One advantage of identifying specific performance issues is that automatically fixing these performance bugs may be achieved by applying code changes which are known to frequently provide a fix \cite{nistor2014understanding}. In the current form of this work, when a performance bug is detected the unit of code marked as relevant to the bug is a (comparitively coarse-grained) function (or method in Java) \cite{nistor2014understanding}.

Coarse-grained approaches which use a method or function as the smallest unit of code considered, appear more scalable for larger programs \cite{mudduluru2016efficient}. We see such approaches as complimentary where progressively less scalable approaches (such as exhaustive mutation) is only used after more scalable approaches have been used to broadly indicate what methods or libraries are associated with a performance issue.

\section{Conclusion \& Future Work}
\label{sec-11}

We have shown how profiling is suited to finding performance bottlenecks and how this differs from locating performance improvements. Mutation analysis can be used to locate performance improvements and we show that even mutations which degrade functionality are worth analysing to locate performance improvements. 

Our approach for using mutation to highlight performance improvements is general in that we do not target any specific type of code nor recommend any type of solution. Our approach could be augmented to utilise many of the known improvement techniques as listed in other work \cite{nistor2015c}. Our approach is designed for automated search and therefore may not be directly beneficial for use by human programmers. Our approach is instead expected to work well with automated program improvement approaches where mutation and testing are performed.
We speculate that it may also be more generally applicable for different types of ``performance''. The basic idea is that by modifying code you can measure changes in the characteristic of interest. There is the potential to apply this approach to find code which has a big impact on memory, network or disk usage provided these characteristics can be measured.   \newpage

\bibliographystyle{IEEEtran}
\bibliography{bibexport}

\begin{thebibliography}{10}
\providecommand{\url}[1]{#1}
\csname url@samestyle\endcsname
\providecommand{\newblock}{\relax}
\providecommand{\bibinfo}[2]{#2}
\providecommand{\BIBentrySTDinterwordspacing}{\spaceskip=0pt\relax}
\providecommand{\BIBentryALTinterwordstretchfactor}{4}
\providecommand{\BIBentryALTinterwordspacing}{\spaceskip=\fontdimen2\font plus
\BIBentryALTinterwordstretchfactor\fontdimen3\font minus
  \fontdimen4\font\relax}
\providecommand{\BIBforeignlanguage}[2]{{%
\expandafter\ifx\csname l@#1\endcsname\relax
\typeout{** WARNING: IEEEtran.bst: No hyphenation pattern has been}%
\typeout{** loaded for the language `#1'. Using the pattern for}%
\typeout{** the default language instead.}%
\else
\language=\csname l@#1\endcsname
\fi
#2}}
\providecommand{\BIBdecl}{\relax}
\BIBdecl

\bibitem{nistor2013discovering}
A.~Nistor, T.~Jiang, and L.~Tan, ``Discovering, reporting, and fixing
  performance bugs,'' in \emph{Proceedings of the 10th Working Conference on
  Mining Software Repositories}.\hskip 1em plus 0.5em minus 0.4em\relax IEEE
  Press, 2013, pp. 237--246.

\bibitem{jones2005empirical}
J.~A. Jones and M.~J. Harrold, ``Empirical evaluation of the tarantula
  automatic fault-localization technique,'' in \emph{IEEE/ACM International
  Conference on Automated Software Engineering}.\hskip 1em plus 0.5em minus
  0.4em\relax ACM, 2005, pp. 273--282.

\bibitem{coppa2012input}
E.~Coppa, C.~Demetrescu, and I.~Finocchi, ``Input-sensitive profiling,'' in
  \emph{ACM SIGPLAN Notices}, vol.~47, no.~6.\hskip 1em plus 0.5em minus
  0.4em\relax ACM, 2012, pp. 89--98.

\bibitem{knuth1974structured}
D.~E. Knuth, ``Structured programming with go to statements,'' \emph{ACM
  Computing Surveys (CSUR)}, vol.~6, no.~4, pp. 261--301, 1974.

\bibitem{selakovic2016performance}
M.~Selakovic and M.~Pradel, ``Performance issues and optimizations in
  javascript: an empirical study,'' in \emph{Proceedings of the 38th
  International Conference on Software Engineering}.\hskip 1em plus 0.5em minus
  0.4em\relax ACM, 2016, pp. 61--72.

\bibitem{mudduluru2016efficient}
R.~Mudduluru and M.~K. Ramanathan, ``Efficient flow profiling for detecting
  performance bugs,'' in \emph{Proceedings of the 25th International Symposium
  on Software Testing and Analysis}.\hskip 1em plus 0.5em minus 0.4em\relax
  ACM, 2016, pp. 413--424.

\bibitem{conf/issta/ShenLPG15}
\BIBentryALTinterwordspacing
D.~Shen, Q.~Luo, D.~Poshyvanyk, and M.~Grechanik, ``Automating performance
  bottleneck detection using search-based application profiling,'' in
  \emph{ISSTA}, M.~Young and T.~Xie, Eds.\hskip 1em plus 0.5em minus
  0.4em\relax ACM, 2015, pp. 270--281. [Online]. Available:
  \url{http://dl.acm.org/citation.cfm?id=2771783}
\BIBentrySTDinterwordspacing

\bibitem{Jin12a}
G.~Jin, L.~Song, X.~Shi, J.~Scherpelz, and S.~Lu, ``Understanding and detecting
  real-world performance bugs,'' in \emph{ACM SIGPLAN Conference on Programming
  Language Design and Implementation (PLDI'12)}.\hskip 1em plus 0.5em minus
  0.4em\relax Beijing, China: ACM Press, Jun. 2012, pp. 77--88.

\bibitem{weimer2009automatically}
W.~Weimer, T.~Nguyen, C.~Le~Goues, and S.~Forrest, ``Automatically finding
  patches using genetic programming,'' in \emph{International Conference on
  Software Engineering (ICSE)}.\hskip 1em plus 0.5em minus 0.4em\relax IEEE
  Computer Society, 2009, pp. 364--374.

\bibitem{langdon2013optimising}
W.~B. Langdon and M.~Harman, ``Optimising existing software with genetic
  programming,'' \emph{IEEE Transactions on Evolutionary Computation}, 2013.

\bibitem{wu2015deep}
F.~Wu, W.~Weimer, M.~Harman, Y.~Jia, and J.~Krinke, ``Deep parameter
  optimisation,'' in \emph{Proceedings of the 2015 Annual Conference on Genetic
  and Evolutionary Computation}.\hskip 1em plus 0.5em minus 0.4em\relax ACM,
  2015, pp. 1375--1382.

\bibitem{langdon:2015:csdc}
\BIBentryALTinterwordspacing
W.~B. Langdon and J.~Petke, ``Software is not fragile,'' in \emph{Complex
  Systems Digital Campus E-conference, CS-DC'15}, ser. Proceedings in
  Complexity, P.~Bourgine and P.~Collet, Eds.\hskip 1em plus 0.5em minus
  0.4em\relax Springer, Sep. 30-Oct. 1 2015, p. Paper ID: 356, invited talk,
  Forthcoming. [Online]. Available: \url{http://cs-dc-15.org/}
\BIBentrySTDinterwordspacing

\bibitem{javaspec}
J.~Gosling, B.~Joy, G.~Steele, G.~Bracha, and A.~Buckley, ``The java language
  specification,''
  \url{https://docs.oracle.com/javase/specs/jls/se7/html/index.html}, Feb.
  2013.

\bibitem{jdt}
{The Eclipse Foundation}, ``Java development tools,''
  \url{http://www.eclipse.org/jdt/}, Nov. 2012.

\bibitem{kuperberg2008a}
M.~Kuperberg, M.~Krogmann, and R.~Reussner, ``{ByCounter: Portable Runtime
  Counting of Bytecode Instructions and Method Invocations},'' in
  \emph{Workshop on Bytecode Semantics, Verification, Analysis and
  Transformation (European Joint Conferences on Theory and Practice of
  Software)}, 2008.

\bibitem{Cody-Kenny:2015:gi}
\BIBentryALTinterwordspacing
B.~Cody-Kenny, E.~G. Lopez, and S.~Barrett, ``{locoGP:} improving performance
  by genetic programming java source code,'' in \emph{Genetic Improvement 2015
  Workshop}, W.~B. Langdon, J.~Petke, and D.~R. White, Eds.\hskip 1em plus
  0.5em minus 0.4em\relax Madrid: ACM, 11-15 Jul. 2015, pp. 811--818. [Online].
  Available: \url{https://www.scss.tcd.ie/~codykenb/locoGP.html}
\BIBentrySTDinterwordspacing

\bibitem{DBLP:journals/corr/Cody-KennyB16}
\BIBentryALTinterwordspacing
B.~Cody{-}Kenny and S.~Barrett, ``Performance localisation,'' \emph{CoRR}, vol.
  abs/1603.01489v1, 2016. [Online]. Available:
  \url{https://arxiv.org/abs/1603.01489v1}
\BIBentrySTDinterwordspacing

\bibitem{jia2011analysis}
Y.~Jia and M.~Harman, ``An analysis and survey of the development of mutation
  testing,'' \emph{IEEE transactions on software engineering}, vol.~37, no.~5,
  pp. 649--678, 2011.

\bibitem{petke2014using}
J.~Petke, M.~Harman, W.~B. Langdon, and W.~Weimer, ``Using genetic improvement
  \& code transplants to specialise a c++ program to a problem class,'' in
  \emph{European Conference on Genetic Programming (EuroGP)}, 2014.

\bibitem{nagaraj2012structured}
K.~Nagaraj, C.~Killian, and J.~Neville, ``Structured comparative analysis of
  systems logs to diagnose performance problems,'' in \emph{Presented as part
  of the 9th USENIX Symposium on Networked Systems Design and Implementation
  (NSDI 12)}, 2012, pp. 353--366.

\bibitem{schulte2012software}
E.~Schulte, Z.~P. Fry, E.~Fast, W.~Weimer, and S.~Forrest, ``Software
  mutational robustness,'' \emph{Genetic Programming and Evolvable Machines},
  vol.~15, no.~3, pp. 281--312, 2012.

\bibitem{harrold1998empirical}
M.~J. Harrold, G.~Rothermel, R.~Wu, and L.~Yi, ``An empirical investigation of
  program spectra,'' \emph{ACM SIGPLAN Notices}, vol.~33, no.~7, pp. 83--90,
  1998.

\bibitem{ball1999concept}
T.~Ball, ``The concept of dynamic analysis,'' in \emph{Software
  Engineering—ESEC/FSE’99}.\hskip 1em plus 0.5em minus 0.4em\relax
  Springer, 1999, pp. 216--234.

\bibitem{conf/icsm/VasquezVLP15}
\BIBentryALTinterwordspacing
M.~L. V{\'a}squez, C.~Vendome, Q.~Luo, and D.~Poshyvanyk, ``How developers
  detect and fix performance bottlenecks in android apps,'' in \emph{ICSME},
  R.~Koschke, J.~Krinke, and M.~P. Robillard, Eds.\hskip 1em plus 0.5em minus
  0.4em\relax IEEE, 2015, pp. 352--361. [Online]. Available:
  \url{http://ieeexplore.ieee.org/xpl/ mostRecentIssue.jsp?punumber=7321954}
\BIBentrySTDinterwordspacing

\bibitem{conf/icse/ManotasPC14}
\BIBentryALTinterwordspacing
I.~L.~M. Guti{\'e}rrez, L.~L. Pollock, and J.~Clause, ``{SEEDS}: a software
  engineer's energy-optimization decision support framework,'' in \emph{ICSE},
  P.~Jalote, L.~C. Briand, and A.~van~der Hoek, Eds.\hskip 1em plus 0.5em minus
  0.4em\relax ACM, 2014, pp. 503--514. [Online]. Available:
  \url{http://dl.acm.org/citation.cfm?id=2568225}
\BIBentrySTDinterwordspacing

\bibitem{grechanik2012automatically}
M.~Grechanik, C.~Fu, and Q.~Xie, ``Automatically finding performance problems
  with feedback-directed learning software testing,'' in \emph{2012 34th
  International Conference on Software Engineering (ICSE)}.\hskip 1em plus
  0.5em minus 0.4em\relax IEEE, 2012, pp. 156--166.

\bibitem{killian2010finding}
C.~Killian, K.~Nagaraj, S.~Pervez, R.~Braud, J.~W. Anderson, and R.~Jhala,
  ``Finding latent performance bugs in systems implementations,'' in
  \emph{Proceedings of the eighteenth ACM SIGSOFT international symposium on
  Foundations of software engineering}.\hskip 1em plus 0.5em minus 0.4em\relax
  ACM, 2010, pp. 17--26.

\bibitem{impagliazzo1998problems}
R.~Impagliazzo, R.~Paturi, and F.~Zane, ``Which problems have strongly
  exponential complexity?'' in \emph{Foundations of Computer Science, 1998.
  Proceedings. 39th Annual Symposium on}.\hskip 1em plus 0.5em minus
  0.4em\relax IEEE, 1998, pp. 653--662.

\bibitem{conf/pldi/OlivoDL15}
\BIBentryALTinterwordspacing
O.~Olivo, I.~Dillig, and C.~Lin, ``Static detection of asymptotic performance
  bugs in collection traversals,'' in \emph{PLDI}, D.~Grove and S.~Blackburn,
  Eds.\hskip 1em plus 0.5em minus 0.4em\relax ACM, 2015, pp. 369--378.
  [Online]. Available: \url{http://dl.acm.org/citation.cfm?id=2737924}
\BIBentrySTDinterwordspacing

\bibitem{nistor2015c}
A.~Nistor, P.-C. Chang, C.~Radoi, and S.~Lu, ``C aramel: detecting and fixing
  performance problems that have non-intrusive fixes,'' in \emph{Proceedings of
  the 37th International Conference on Software Engineering-Volume 1}.\hskip
  1em plus 0.5em minus 0.4em\relax IEEE Press, 2015, pp. 902--912.

\bibitem{nistor2014understanding}
A.~Nistor, ``Understanding, detecting, and repairing performance bugs,'' Ph.D.
  dissertation, University of Illinois at Urbana-Champaign, 2014.

\end{thebibliography}
\end{document}